\documentclass{article}
\usepackage{smc2026}
\usepackage[caption=false, font=footnotesize]{subfig}
\usepackage{paralist}
\usepackage[figure,table]{hypcap}

\usepackage[english]{babel}
\usepackage[whole]{bxcjkjatype}
\usepackage{adjustbox}
\usepackage{algorithm}
\usepackage{algpseudocode}
\usepackage{stfloats}

\def\papertitle{Score-Agnostic Structure Analysis \\in Large-Scale Performance Datasets}

\author[1]{\mbox{\firstname{Patricia}\lastname{Hu}\originalname{胡紫漪}}}
\author[1, 2]{\mbox{\firstname{Silvan}\middlename{David}\lastname{Peter}}}
\author[1, 2]{\mbox{\firstname{Gerhard}\lastname{Widmer}}}

\affil[1]{\institution{Institute of Computational Perception, Johannes Kepler University}\city{Linz}\country{Austria}\affiliationtype{University}}
\affil[2]{\institution{LIT AI Lab, Linz Institute of Technology}\city{Linz}\country{Austria}\affiliationtype{Music}}

\completesetup

\title{\papertitle}
\begin{document}
	\capstartfalse
	\maketitle
	\capstarttrue

\section{Introduction}\label{sec:introduction}
Music performance research involves understanding and interpreting the nuances of expressive musical renditions. To date, most work in expressive performance analysis has relied on small, curated datasets and/or manual annotation. In the case of solo piano music, most work has relied on MIDI-like data measured on computer-controllable instruments.
In recent years, thanks to advances in automatic music transcription (AMT) \cite{kong2021highresolution}, several large-scale datasets of automatically transcribed piano solo music have been released \cite{kong2020giantmidi, zhang2023atepp, bradshaw2025aria}. While these datasets undoubtedly offer extensive material for performance studies, they vary substantially in quality. Their curation workflows rely primarily on automated techniques, i.e., using an audio classifier to gather and label audio recordings with relevant metadata such as composition and performer, and subsequently converting those labelled recordings into a symbolic representation using AMT models. 
In the case of classical music, performances often differ not only in expressive aspects such as tempo, but also in their structural interpretation of the score (including repeat patterns and edition-specific variants). To meaningfully use large-scale transcribed datasets for performance research, transcriptions of the same piece must be grouped according to their underlying structural realisation to support valid comparison.
We address this by applying sequence-to-sequence alignment followed by hierarchical clustering: we create pairwise alignments for all pairs of transcriptions of a given piece, and use the alignment cost and (dis)similarity of performed sequence lengths to resolve structural mismatches as features for grouping. We propose this approach as a first step towards automatically evaluating large-scale transcribed datasets.


\section{Method}

Our proposed method works in two steps: first, we compute alignments of all pairs of transcriptions of a given piece using Dynamic Time Warping (DTW) with a custom distance metric. Second, we cluster the transcriptions based on their alignment cost, stretch (warping) and sequence length (dis)similarity.

\subsection{Sequence-to-sequence alignment}
Given a list of transcriptions $\mathcal{T} = [T_1, T_2, \ldots, T_n]$, where $T_i$ is a sequence of notes, we first convert each transcription $T_i$ into a list of chords $C_i$, yielding a corresponding set $\mathcal{C} = [C_1, C_2, \ldots, C_n]$ of chordwise performance representations.

To do so, we group all notes of a given transcription into a sequence of chords using two parameters: an inter-onset-interval (IOI) threshold $\tau_{\text{IOI}}$ and a chord duration threshold $\tau_{\text{chord}}$. The parameter $\tau_{\text{IOI}}$ determines the maximum time gap between notes belonging to the same chord, whereas $\tau_{\text{chord}}$ specifies the maximum difference in onset times between the first and last note in a chord, both measured in seconds. We define the onset of a chord event as the mean onset time of all its constituent notes. Next, we normalize all chord onset times over the entire sequence, yielding a relative time representation suitable for the $\text{cost}_{\text{time}}$ component of the alignment cost function described below. The pitches of each chord are encoded by their respective pitch classes.

Next, we form all pairs $\{C_i, C_j\}$ for $1 \leq i < j \leq n$ of those chord sequences and apply Dynamic Time Warping (DTW) \cite{sakoe1978dynamic, peter2023automatic} to find an alignment path between each pair. We define a custom distance metric between chords $c_i \in C_i$ and $c_j \in C_j$ that balances harmonic similarity and timing differences:

\begin{align}
    \text{cost}(c_i, c_j) = \; & \alpha \cdot \text{cost}_{\text{pitch}}(c_i, c_j) \nonumber \\
                               & + (1 - \alpha) \cdot \text{cost}_{\text{time}}(c_i, c_j)
\end{align}

\noindent where $\text{cost}_{\text{pitch}}$ uses the Jaccard distance for harmonic comparison, $\text{cost}_{\text{time}}$ is the absolute difference in normalised chord onset times, and $\alpha$ is a weighting factor balancing the two components. For each pair of transcribed performances, this yields an alignment path along with its cumulative cost.

\subsection{Hierarchical Clustering}

\begin{figure*}[!t]
    \centering
    {\includegraphics[width=2\columnwidth]{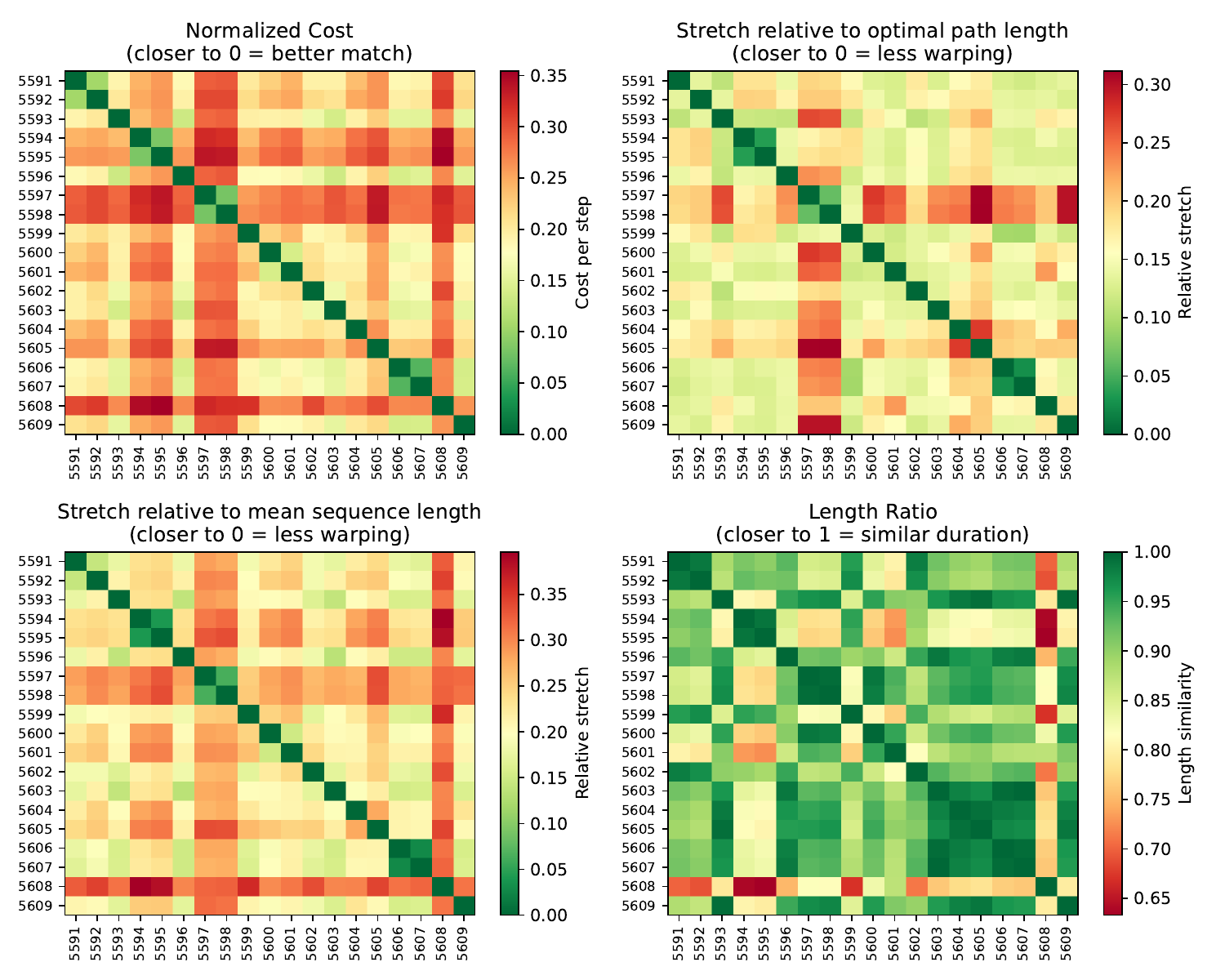}} 
    \caption{Four distance matrices derived from all pairwise alignments of transcriptions from the 3rd movement of the Piano Sonata KV 331 by W. A. Mozart (taken from the ATEPP dataset). These capture: (1) alignment cost, (2) warping relative to optimal path length, (3) warping relative to mean sequence length, and (4) sequence length ratio. Axis labels denote the transcription IDs as labelled in the dataset.}
    \label{fig:features}
\end{figure*}

Given the alignment path and its cost, along with the sequence lengths $I$ and $J$ for each pair $\{C_i, C_j\}$,, we build four distance matrices that capture normalised alignment cost, temporal warping relative to the optimal path, temporal warping relative to the average chord sequence length, and the ratio of the two sequence lengths. 
See \ref{fig:features} for an illustration of these distance matrices.
We then use a weighted combination of the four distance matrices as the input feature for hierarchical clustering \cite{mullner2011modern}.

\section{Demonstration}

We demonstrate our method on the ATEPP dataset  \cite{zhang2023atepp}, which contains 11,674 transcriptions of performances by world-renowned pianists of 1,595 unique compositions, for 319 (20\%) of which score files are provided. 
We test our method on all transcriptions of works by Haydn, Mozart, Beethoven, Schubert and Schumann which provide a corresponding score file, and contain more than one transcription and structural realisation. In total, this covers 88 compositions and 1,516 transcriptions. We use the score-dependent repeat estimator from \cite{peter2025infer} as a baseline and evaluate our score-agnostic structural clustering approach on homogeneity, completeness and V-Measure \cite{rosenberg2007v}. Homogeneity measures whether each cluster contains items from a single group, completeness captures whether all items belonging to one group are clustered together, and V-Measure is their harmonic mean.

We run a grid search on 77 (of the 88) pieces (1.220 transcriptions) to find the (distance matrix) weights, cluster method and distance threshold that approximate the estimations produced by the score-dependent repeat identifier. Generally, we find that to maximise cluster homogeneity, the cost matrix and the amount of warping relative to the optimal path play a more significant role than the warping relative to the average sequence length or the ratio between sequence lengths, both of which are weighted higher when optimising for cluster completeness.
We prioritize homogeneity over completeness in the clustering to avoid the spurious inclusion of mismatching performances, so we report results using parameters optimised for homogeneity. On the unseen 11 pieces (296 transcriptions), we achieve a mean homogeneity score of 61.05\%. Given that the baseline score-dependent structure identifier is an estimator itself, we manually verified and corrected the structure labels for the 296 transcriptions corresponding to the 11 pieces, after which we achieve a mean homogeneity score of 96.39\%.

\begin{figure*}[!t]
    \centering
    {\includegraphics[width=1.8\columnwidth]{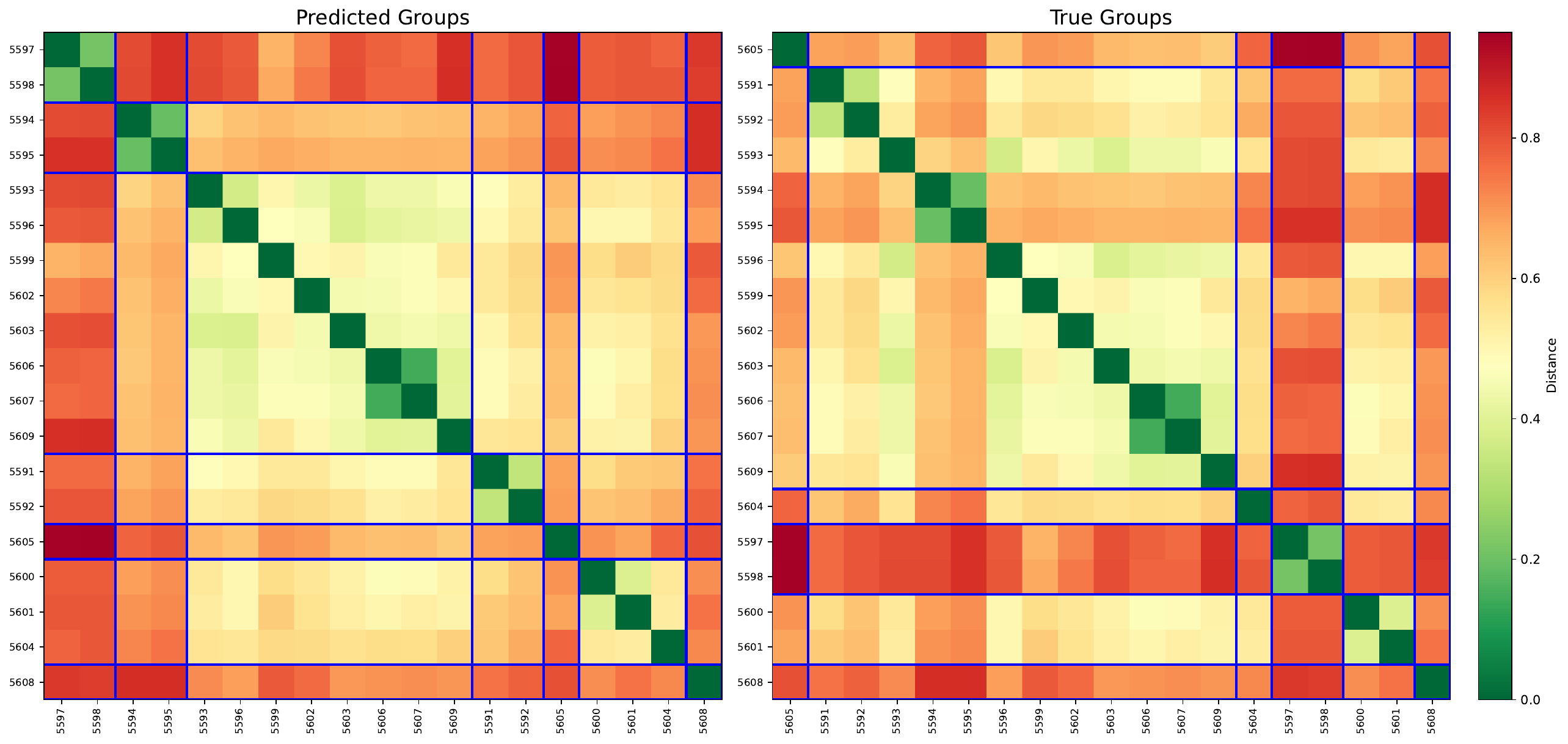}} 
    \caption{A comparison of the estimated grouping by our score-agnostic clustering (left) and the true (manually verified) groupings (right), both for the same piece and transcriptions as in \ref{fig:features}. The underlying feature matrix is a weighted combination of the four distance matrices shown there. Note that the transcription IDs are ordered differently in both matrices to create contiguous groupings (blue lines) for visualisation purposes.
    }
    \label{fig:true_vs_predict}
\end{figure*}

\begin{figure*}[!b]
    \centering
    {\includegraphics[width=1.7\columnwidth]{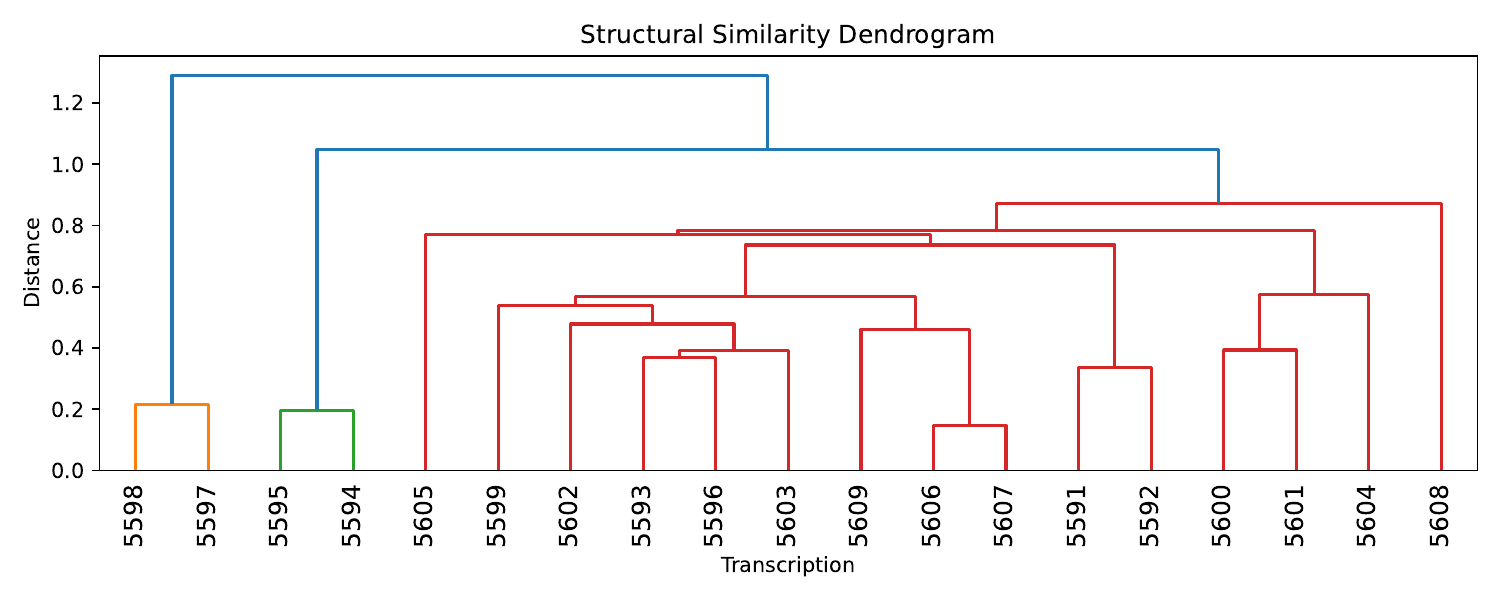}} 
    \caption{A dendrogram visualisation of the hierarchical clustering process as a tree-like diagram (same piece and transcriptions as in \ref{fig:features}).
    }
    \label{fig:dendrogram}
\end{figure*}

\begin{figure*}[!b]
    \centering
    {\includegraphics[width=1.7\columnwidth]{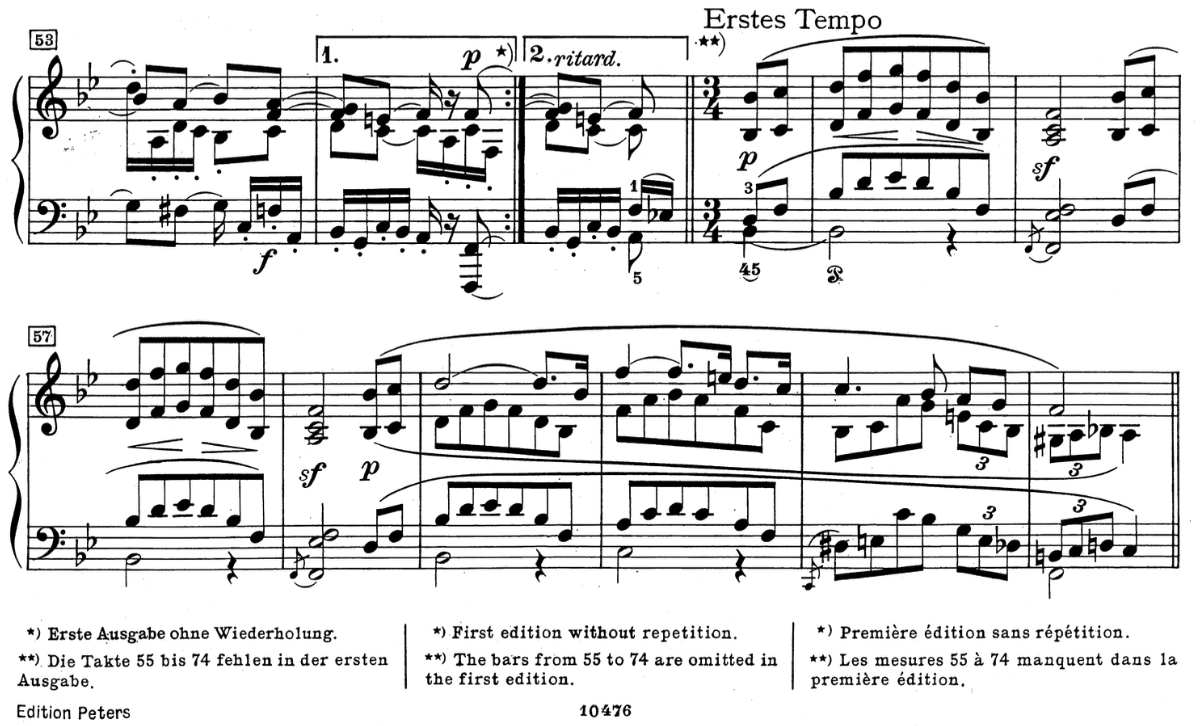}} 
    \caption{Score image excerpt taken from the second movement of the Kreisleriana Op. 16 by R. Schumann from the Peters Edition. The composition exists in different notated versions (see footnote ** in the image), but only one of these is reflected in the digital score representation in the dataset.
    }
    \label{fig:score_edition}
\end{figure*}

Fig. \ref{fig:true_vs_predict} shows a comparison of our score-agnostic grouping and the true performed group structures. Fig. \ref{fig:dendrogram} visualises the hierarchical clustering process for the given transcriptions of this piece as a tree-like diagram, illustrating how similar transcriptions are progressively grouped.
We notice that our score-agnostic grouping is more robust to structural differences resulting from both encoding errors (e.g., missing repeat signs) and varying editions, compared to the score-dependent repeat identifier. 
An example of the latter is Fig.~\ref{fig:score_edition}, which shows an excerpt from the score of Schumann’s Kreisleriana Op. 16, second movement. After manually verifying the grouping structure estimated by the score-dependent repeat identifier for this piece, we found that the score exists in various editions, which are not captured by the digitally encoded score edition in the dataset, yet our algorithm correctly grouped the performed structures.

Additional challenges for any structure grouping algorithm come from artifacts that are introduced by the audio transcription (e.g., missed or wrongly inserted notes, anomalous rhythmic binning of note onset times, etc.), depending on the quality of the source audio. Compared to the score-dependent repeat identifier, we noticed that our score-agnostic grouping is more robust to transcriptions with pitch-related artifacts, and tends to isolate those with rhythm-related artifacts into separate groups.

\section{Conclusion}

We presented an approach to automatically align and cluster transcriptions of a given composition into their different structural score realisations. We propose this approach as a first step towards automatically evaluating transcribed datasets that lack ground-truth score and/or audio, shifting the evaluation criterion from truth-based accuracy to musical coherence and plausibility. As large-scale symbolic music datasets grow in size and diversity, manual quality control becomes infeasible. Our approach offers a scalable, reference-free means of assessing transcription quality for the curation and maintenance of encoded music collections.

Our method is available in the python piano transcription library mpteval:
\url{https://github.com/CPJKU/mpteval}.
Additionally the demonstration can be found here:
\url{https://github.com/huispaty/score-agnostic-structuring}.

\begin{acknowledgments}
This research acknowledges support by the European Research Council (ERC), under the European Union’s Horizon 2020 research and innovation programme, grant agreement No. 101019375 Whither Music?.
\end{acknowledgments}

\bibliography{refs}
	
\end{document}